\documentstyle[epsfig,amstex,amssymb]{mn}

\newcommand{\ie}{i.e.\ }
\newcommand{\eg}{e.g.\ }
\newcommand{\lya}{Ly$\alpha$\ }
\newcommand{\civ}{\hbox{C\,{\sc iv}}}
\newcommand{\ciii}{\hbox{C\,{\sc iii}]}}
\newcommand{\mgii}{\hbox{Mg\,{\sc ii}}}

\newcommand{\oiii}{\hbox{[O\,{\sc iii}]}}
\newcommand{\hb}{H$\beta$}
\newcommand{\ha}{H$\alpha$}
\newcommand{\feii}{\hbox{Fe\,{\sc ii}}}

\title{Red synchrotron jets in Parkes quasars}  
\author[M.T.Whiting, R.L.Webster and P.J.Francis] {M.T.Whiting$^1$, 
R.L.Webster$^1$ and P.J.Francis$^{2,3}$ \\ 
$^1$ Astrophysics group, School of Physics, University of Melbourne,
Victoria 3010, Australia\\  
$^2$ Research School of Astronomy and Astrophysics, Australian
National University, Canberra, ACT 0200, Australia\\ 
$^3$ Joint Appointment with the Department of Physics and Theoretical
Physics, Faculty of Science}

\begin{document}

\maketitle

\begin{abstract}
We present model fits to spectral energy distributions in the optical
and NIR of $>100$ flat-spectrum radio quasars from the Parkes
Half-Jansky Flat-spectrum Sample. We find that $\sim 40\%$ of the
sources have power-law SEDs, while a similar number show evidence for
two primary components: a blue power law and optical synchrotron
emission. The blue power law is similar to the dominant component
observed in the spectra of optically-selected QSOs. There is strong
evidence that the synchrotron component has a turnover in the
UV-optical rest frame of the spectrum. In the remaining sources it is
likely that the synchrotron peaks at longer wavelengths. This mixture
of two components is supported by optical polarisation measurements in
a subgroup of the sources. The sources with power law SEDs show
evidence for an excess number of red power law slopes compared to
optically-selected quasars. There are additional spectral components
in some of the sources, such as dust and the underlying galaxy, which
have not been considered here.
\end{abstract}


\section{Introduction}
\label{sec-intro}

The spectral energy distributions of Active Galactic Nuclei provide a
sum of the emission processes contributing to the energy output of the
AGN.  Clearly, several mechanisms contribute at most frequencies.  If
we can determine the different components as a function of frequency,
then we can determine the energy generation mechanisms which are
important.

Flat-spectrum radio quasars are known to be dominated in the radio
spectrum by synchrotron emission. It is less clear how far these
synchrotron spectra extend towards higher frequencies, or the
magnitude of the contribution they make to the optical and
near-infrared (NIR) spectrum -- a region dominated by the Big Blue
Bump.

The far-IR emission of flat-spectrum quasars appears to be dominated by
synchrotron emission \cite{haas98}, consistent with a Doppler-boosted
component that swamps the expected emission from dust components.
Synchrotron emission in the optical has been considered for many years
to be necessary to explain the optical to NIR photometry of optically
faint and red flat-spectrum radio quasars. This has mostly been on the
basis of the steepness of the optical--NIR spectra
\cite{rieke79,beichmann81,bregman81,rieke82}. The steepness of these
spectra suggested a cutoff, or sharp break, in the electron energy
spectrum. Supporting evidence for the presence of synchrotron emission
at NIR wavelengths comes from both the variability in the NIR, and
from the strong polarisation of a few sources. Stickel et al.\
\shortcite{stickel96} found that variability was a general
characteristic of optically faint flat-spectrum radio sources,
although they did not specifically discuss synchrotron
emission. Instead, they suggest that reddening due to intervening
galaxies or the host galaxy itself was the likely cause of the red
optical colours.

While radio jets are quite common, and have been widely observed,
optical counterparts to these jets are quite rare. In fact, only 14
optical jets are currently known (see O'Dea et al.\ \shortcite{odea99}
and references therein). The features (\ie emission knots) in the
optical jets match the positions of features in the radio jets,
indicating that the emission regions and mechanisms for the two
spectral regimes are associated. Optical jets can be used to put
strong constraints on the particle energetics, by providing limits on
the maximum energies, and on the acceleration mechanisms, due to the
short lifetimes of particles at these energies \cite{meisenheimer96}.

The quasars considered in this paper are selected from the Parkes
Half-Jansky Flat-spectrum Sample (PHFS) \cite{drinkwater97}, which
consists of 323 objects selected to be radio-loud (${\rm S}_{2.7\ {\rm
GHz}}>0.5\ {\rm Jy}$), and have a flat radio spectrum
($\alpha_{2.7/5.0}>-0.5$, $S(\nu)\propto \nu^\alpha$). These quasars
have been shown \cite{webster95,francis00} to have a large spread in
$B-K$ colours, with the reddest objects having $B-K>7$. Masci, Webster
\& Francis \shortcite{masci98} showed that this spread could not be
accounted for by the emission from the host-galaxy.

In this paper we test the idea that optical synchrotron emission
causes the red optical-NIR colours of the PHFS quasars.  Models
representing emission from both an optical synchrotron component and a
blue optical power law (representing the continuum emission from an
unreddened quasar) are fitted to broad-band optical-NIR spectra.
Standard goodness-of-fit techniques are used to determine models which
are consistent with the observations.

The dataset was compiled by Francis, Whiting \& Webster
\shortcite{francis00} (FWW hereafter), and comprises broadband optical
and NIR photometry in the bands $B$, $V$, $R$, $I$, $J$, $H$ and
$K$. This photometry is quasi-simultaneous, meaning all observations
for a given source were made within several days (six at most) of one
another.  This minimises the effects of source variability. The source
selection used for this paper is explained in Section~\ref{sec-data}.
The simultaneity, as well as the breadth and density of the spectral
coverage of this dataset provides an excellent basis to model the
broadband emission from a large number of flat-spectrum quasars.

The dataset is described in Section~\ref{sec-data}. In
Section~\ref{sec-powerlaw} we test the hypothesis that the emission in
the optical-NIR region is well-fitted by a single power law.  In
Section~\ref{sec-models} we describe the more sophisticated model that
was fitted to the data -- representing the accretion disk and
synchrotron emission -- and the method that was used to perform the
fitting.  The results of this fitting are described in
Sections~\ref{sec-results} \& \ref{sec-synchprops}, while further
tests of the model fits are examined in Section~\ref{sec-tests}, using
polarisation and emission line measurements. A possible rival model --
that of black-body emission by hot dust -- is considered in
Section~\ref{sec-dust}. The effect of emission lines on the photometry
and the resulting fits is investigated in
Section~\ref{sec-em-effect}. Finally, Section~\ref{sec-discuss}
contains discussion of the results and their implications.

\section{Data}
\label{sec-data}

The data which have been fitted by these models are described in
detail in FWW. A total of 157 sources from the PHFS were observed,
with quasi-simultaneous broadband photometry observations in the bands
$B$, $V$, $R$, $I$, $J$, $H$ and $K$. These magnitudes were converted
into broadband fluxes using the zero points given in the same paper.

\subsection{Excluded sources}

Not all sources in FWW were used. Firstly, we only considered those
sources with complete contemporaneous photometry in all bands, or
those with only one observation missing (either not observed or an
upper limit). This was to ensure that the number of degrees of freedom
in the model fitting was greater than zero. Those sources with more
than one band missing were not included in the analysis. Sources
without a measured redshift were also excluded, since the redshift is
needed to obtain the correct shape of the observed synchrotron
spectrum. One of the sources that had an unknown redshift ($z=0$) in
Drinkwater et al.\ \shortcite{drinkwater97}, 0829$+$046, has a
published redshift value of $z=0.18$ \cite{falomo91}, which is used
here. Also excluded were low redshift galaxies, which had a prominent
4000\AA\ break between the $B$ and $V$ bands. These sources show
strong evidence for flux from the underlying galaxy in their spectra
\cite{masci98}, and thus need an additional galactic component to be
modelled accurately.  Sources with $z>3$ were also excluded, as these
had strong Ly$\alpha$ breaks present between $B$ and $V$. This reduced
the number of sources from 157 to 117.

Twenty-one of the reddest sources ($B-K > 5$) in the subsample were
amongst those excluded, as they have only upper limits on $B$, $V$ and
possibly $R$, or no redshift.  The sources with upper limits are both
red and optically faint. Since the flux of these sources typically
decreases rapidly in the blue (\ie from $V$ to $B$), there are three
possible explanations for them. They are either dust-reddened (causing
the blue decrease in flux), high redshift (and hence \lya absorbed),
or they are dominated by synchrotron that turns over rapidly. Future
analysis of the spectral energy distributions (SEDs) of these sources
will only strengthen our final conclusions.

Those sources excluded purely because of their lack of measured
redshift generally have high $B-K$ colours ($B-K>5$), since the most
likely reason they have no redshift is that they are faint in the
optical (and in particular in the blue).  Further discussion of the 
sources excluded from the sample is given in Section~\ref{sec-discuss}.

\subsection{Errors on photometry}
\label{sec-errphot}

The photometry given in FWW quoted error bars, where the estimated
error comprised two parts: a random error component and an assumed 5\%
error in the photometric zero points, which were added together in
quadrature. The photometric zero point errors were estimated from the
scatter in zero points between different standard star measurements in
an individual night: Francis et al.\ adopted a value of 5\% to account
for this error.

However, this zero point error ignores a number of factors that we
believe may be important for the analysis in this paper. These
factors are as follows:
\begin{itemize}
\item The optical zero magnitude fluxes were taken from Bessell,
Castelli \& Plez \shortcite{bessell98}. These fluxes were derived for
an A0 star, and so will be slightly incorrect for a quasar
spectrum. This has the effect of introducing small colour terms into
the photometry, the size of which will depend on the spectral index of
the object being observed. A similar effect will of course be present
in the NIR. Bersanelli, Bouchet \& Falomo \shortcite{bersanelli91}
found that spectral shape differences could produce systematic errors
of at least a few percent in the NIR flux.
\item The zero point fluxes in the optical are taken from a different
reference \cite{bessell98} to those in the NIR, which were calculated
by P.McGregor ({\it CASPIR} manual, MSSSO, ANU), assuming that Vega is
well represented in the NIR by a black body of temperature 11200 K,
and normalisation $F_\lambda(555 \rm{nm}) = 3.44 \times 10^{-12}
\rm{W\ cm}^{-2}\ \mu {\rm m}^{-1}$ \cite{bersanelli91}.  These
different zero points may not be exactly equivalent, which will
produce small offsets between the optical and NIR parts of the SED.
\item If the sky conditions at Siding Spring Observatory were not
completely photometric for all observations for a given source
(particularly if the transparency changed between different
bands), then the measured photometry will have small band-to-band
errors present.
\item The observations for each source were taken
quasi-simultaneously (meaning all observations were made within at
most a six-day period), to minimise the effects of
variability. However, a number of the sources in the PHFS have been
found to exhibit intra-day variability in the optical
\cite{romero99,heidt96}, and a larger number no doubt have similar
properties to these. Therefore, variability on timescales of the order
of those separating our observations is likely for some of the
sources. Such variability can be up to 0.1 mag over the period of a
night. 
\item Finally, the presence of strong emission lines in the quasars'
spectra could boost the flux of a band above the level of the
continuum. We have tried to quantify this effect in
Section~\ref{sec-em-effect}, although this is not possible to do for
many sources, due to the lack of both photometry and a spectrum. Some
additional effects may be due to line blends (such as \feii\ blends),
but this is probably not as strong an effect as you would find in a
radio-quiet sample. 
\end{itemize}

Hence, to take account of all these factors, we have increased the
systematic error in the photometry to 10\%. The random error is kept
at the same level as that presented in FWW, and is added in quadrature
to the systematic error.

\section{Power law model}
\label{sec-powerlaw}

The object of our analysis is to find physical models to explain the
optical and NIR emission. FWW found that about 90\% of the PHFS have
approximately power law SEDs. We first wish to test this more
rigorously. As a starting point, we choose to fit (naively) a simple
power law, with an unconstrained spectral index. This will separate
out the sources that have power law SEDs from those that show some
curvature in their spectrum.

\subsection{Fits to data}

The model we choose to fit is $f_{PL}(\lambda) = c \lambda^\alpha$, so
that the normalisation $c$ and the spectral index $\alpha$ vary. (Note
that this model implies $f_\nu \propto \nu^{-2-\alpha}$.)

This model is fit to the data using a least-squares method. This
generates a $\chi^2$ value: 
\[ \chi^2 = \sum_{i=1}^7 \frac{(y_i-f_{PL}(\lambda_i))^2}{\sigma_i^2}, \] 
which indicates the goodness of fit. A fit to a source will be deemed
to be ``good'' when the value of $\chi^2$ is less than the cut-off
value corresponding to the 99\% confidence level of the $\chi^2$
distribution. For a source with 5 degrees of freedom (as is the
case for most sources with this power law model), this cutoff level is
15.09. (Note that increasing the value of this cutoff is equivalent to
increasing the confidence level -- for example, the 99.5\% cutoff is
16.75 for 5 degrees of freedom.) If the $\chi^2$ value is greater than
this cutoff level, then we reject the null hypothesis that the power
law model fits the data.

\begin{figure}
\epsfbox{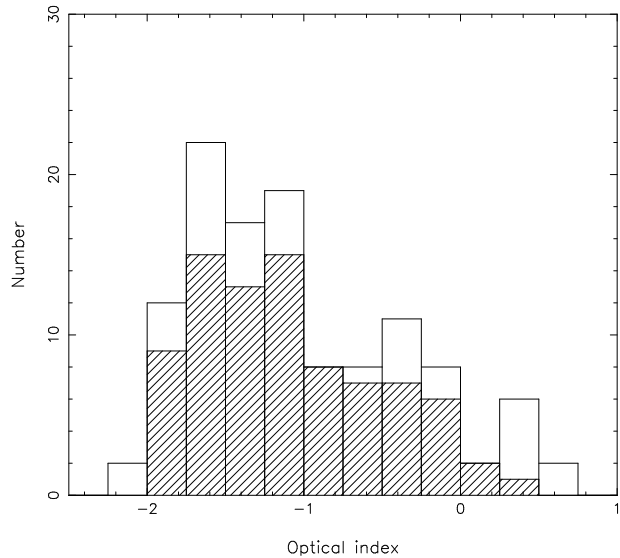}
\caption{Histogram of fitted power law indices. Hatched histogram
indicates good fits, while open histogram indicates all fits.}
\label{fig-index-xray}
\end{figure}

When we fit this power law model to the data, we find that 83 sources
(or 71\% of the total) have good fits. The distribution of resulting
power law indices, both for the good fits and for all sources, is
shown in Fig.~\ref{fig-index-xray}.

The spectral indices of these good fits span a wide range of
values. At one extreme there are the sources with relatively blue SEDs
($\alpha \lesssim -1.4$). These sources are characterised by their
blue continuum, the presence of moderate to strong emission lines, and
generally low X-ray flux (most were not detected by {\it ROSAT}
\cite{siebert98}).

At the other extreme are the sources with redder SEDs (that is,
flatter in $f_\lambda$), with $-1\lesssim\alpha\lesssim0$. These
sources are blazar type objects, with high optical polarisation
\cite{wills92} as well as relatively weak (or even absent) emission
lines -- in fact, all the BL Lac objects in our sample are in this
region.

\subsection{Interpretation of power laws}

So, we have fit a power law to a large majority of the sources in our
sample, spanning a wide range of spectral indices. Are the physical
processes that generate this power law the same for all sources? That
is, does the power law in the blue sources have the same origin as
that in the red sources?

The first class of sources -- the blue sources -- are being fit by a
blue power law, which has similar colours to the blue power law emission
seen in optically-selected quasars \cite{francis96}. This is likely to
be the optical part of the continuum emission from the accretion disk
(often termed the big blue bump).

However, the power law being fit to the redder sources is most likely
of different origin to that seen in the blue sources. These objects
exhibit characteristics commonly associated with optical synchrotron
emission (such as high optical polarisation and lack of prominent
emission lines), and so we postulate that this emission is, at least
in part, some form of synchrotron emission. The slope of the power law
can then be used to determine $p$, the power law index of the electron
energy distribution (\ie defined such that $N(E)\propto E^{-p}$: see
Section~\ref{sec-synch}). Using the values shown in
Fig.~\ref{fig-index-xray}, we obtain $2<p<6$ (using $-1.5<\alpha<0.5$
and $p=2\alpha+5$).

If synchrotron emission is present in the spectra of at least some of
these quasars, then we can ask the question ``Is the synchrotron
component best modelled by this power law?'' The synchrotron component
will be present in one of two forms: a power law caused by an unbroken
(power law) electron energy distribution, or a turning-over component
caused by a break or a cut-off in the electron energy
distribution. (Note that the power law can also be produced by a
synchrotron spectrum turning over at higher frequencies than those
observed.)

Both of these forms can be tested. The power law model must produce
power law indices that are consistent with slopes of plausible energy
distributions. The presence or otherwise of a turnover can be
evaluated by examining the sharpest possible turnover (caused by an
abrupt cut-off in the energy distribution at some maximum
energy). This will provide the maximum contrast with the power law,
and is consistent with modeling done by other authors \cite[for
example]{meisenheimer96}.

Many of the sources, while they have power law fits that we can not
reject at the 99\% confidence level, show evidence for curvature in
their SEDs. This curvature can be `n'-shaped (higher in the middle
than at either end), `u'-shaped (lower in the middle) or perhaps take
the form of an inflection (\eg the flux decreases, levels off and
decreases again). (See Fig.~\ref{fig-ex} for illustrations of the
different types.)  Most of the sources classed as BL Lac objects are
`n'-shaped, and so we postulate that these sources are dominated by a
synchrotron component that is turning over in the optical.

Other sources, however, are bluer in the optical than in the NIR (\ie
show an inflection, without the turn-up seen in `u'-shaped
sources). This is a possible indication of the presence of excess
emission in the NIR, in addition to a blue power law. We propose that
this excess emission is due to a synchrotron component that has turned
over in the NIR, and so does not dominate in the optical, where the
dominant emission is instead a blue power law similar to that observed
in the bluest sources.

\section{Physical models}
\label{sec-models}

In response to this phenomenological classification, we propose the
following physical model. There are two components in this model: one
is a blue power law, representing continuum emission from the
accretion disk region; and the second is synchrotron emission,
representing emission from the relativistic jet (that we know to be
present due to the flat-spectrum radio emission seen in these
objects).

\subsection{Accretion Disk emission}
\label{sec-bbb}

We find, from the simple power law fitting, that the bluest sources
have power law continua. The slopes of these power laws are consistent
with them being the same component as that seen in optically-selected
quasars, commonly termed the ``Big Blue Bump'' (BBB).
We take this component to represent the underlying quasar continuum in
the UV-optical part of the rest-frame spectrum -- that is, the
``un-reddened'' quasar continuum.

Over the wavelength of our observations, the BBB is modelled as a
simple power law, $f_\lambda\propto \lambda^{\alpha_B}$ or $f_\nu
\propto \nu^{-2-\alpha_B}$. Francis \shortcite{francis96} found the
median slope for a subsample of quasars from the Large Bright Quasar
Survey (LBQS), taken from optical/NIR photometry, was $f_\nu \propto
\nu^{-0.35\pm0.3}$, and noted that the observations were consistent
with an intrinsic continuum slope of $f_\nu \propto \nu^{-0.3}$ that
is reddened by various amounts of dust. We therefore take our value of
$\alpha_B$ to be $-1.7$. The effects of allowing the value of
$\alpha_B$ to vary are considered in Section~\ref{sec-discuss}.

\subsection{Synchrotron emission}
\label{sec-synch}

All the PHFS sources are radio-loud flat-spectrum sources, and thus
very likely have relativistic jets that emit synchrotron radiation, at
least at radio frequencies. Could this synchrotron emission extend up
to the optical/NIR part of the spectrum? Our power law fitting from
the previous section provides circumstantial evidence for this: the
redder sources in the optical tend to be the ones with higher
polarisation (a good sign of synchrotron emission) and less prominent
emission lines (possibly a sign that the emission lines are being
swamped by the presence of a synchrotron component). These pieces of
evidence are investigated more deeply in Section~\ref{sec-tests}.

As discussed above, the synchrotron spectrum could take the form of
either a power law, from a power law distribution of electron
energies, or a power law with a break or turn-over, due to an electron
energy distribution that exhibits a break or even a cut-off. This
latter type of spectrum has been seen in optical synchrotron jets
\cite{odea99,scarpa99a}, where the optical spectrum is like
$\nu^{-1.2}$--$\nu^{-3.0}$, compared to a radio--optical spectrum of
$\nu^{-0.6}$--$\nu^{-1.0}$. 

Additionally, a synchrotron spectrum that has a turn-over will, when
combined with the blue power law, be able to reproduce an
inflection-like SED. Such a spectrum, particularly in the region of
the turn-over, will also be quite red, thus accounting for the red
colours of many of the SEDs.

\subsubsection{Analytic modelling}

We consider here synchrotron emission from a population of electrons
with an energy distribution with the form of a power law up to some
maximum energy and zero beyond this (\ie an energy spectrum with an
abrupt cutoff). This can be expressed as a distribution of the Lorentz
factor $\gamma$ of the radiating electrons:
\[ 
N(\gamma)\:{\rm d}\gamma\propto
		\begin{cases}
		\kappa\gamma^{-p}\:{\rm d}\gamma&1\leq\gamma\leq\gamma_c\\
		0				&\gamma>\gamma_c
		\end{cases} 
\]

This is consistent with modelling done by Meisenheimer et al.\
\shortcite{meisenheimer96} on the jet of M87. They found that the
overall synchrotron spectrum of the brightest parts of the jet was
best described by a spectrum that had a sharp cutoff at $\nu_c \approx
10^{15}\, {\text Hz}$, with an energy distribution of the form of a
straight power law $N(\gamma)\propto\gamma^{-2.31}$, with a rather
abrupt high energy cutoff. We consider the effect of using a power law
synchrotron spectrum instead in Section~\ref{sec-discuss}.

\begin{figure*}
\epsfbox{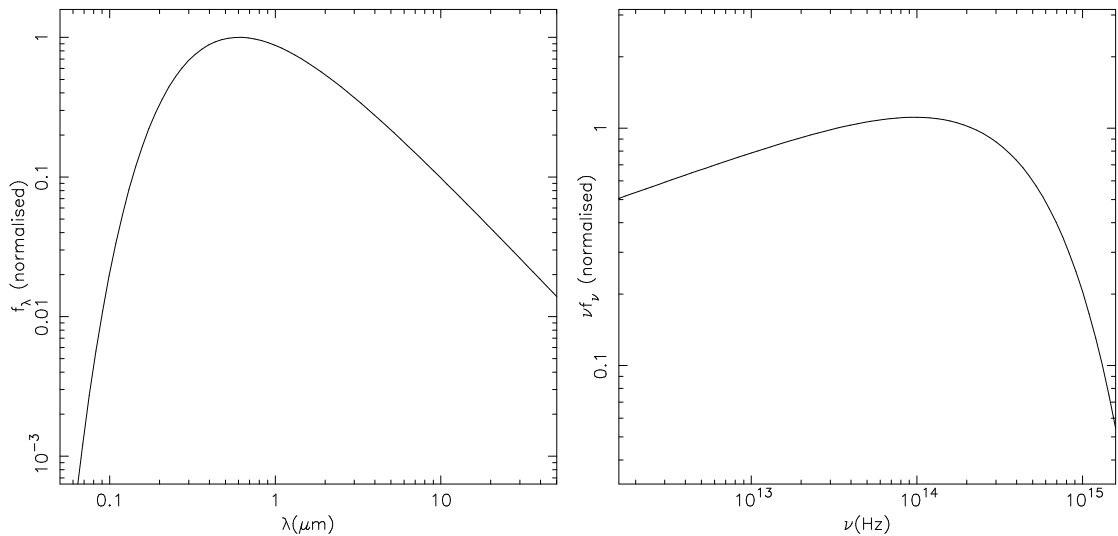}
\caption{The synchrotron model used in the analysis, in both
$f_\lambda$ and $\nu f_\nu$ units. Both plots have been normalised so
that the peak flux in $f_\lambda$ is 1. Note that the vertical scale
is different in the two plots.}
\label{fig-synch}
\end{figure*}

Such a synchrotron spectrum is straightforward to model
analytically. We use the ``classical'' synchrotron model, as first
calculated by Schwinger \shortcite{schwinger49}, and as derived by a
number of authors, particularly Pacholczyk \shortcite{pacholczyk70}
and Longair \shortcite{longair94}, whose derivation we follow.

The single particle luminosity for a radiating electron (mass $m_e$,
charge $e$, Lorentz factor $\gamma=E/m_e c^2$, and pitch angle, or
angle between trajectory and magnetic field direction, $\theta$) in a
uniform magnetic field $B$ is given by
\[ L_{\rm{1p}}(\omega) =
\frac{\sqrt{3}e^3B\sin\theta}{8\pi^2\epsilon_0cm_e} F(x) \]
where $\omega=2\pi\nu$ is the angular frequency,
\[ x = \frac{2\omega\beta m_e}{3\gamma^2eB\sin\theta}, \]
and $F(x)$ is defined in terms of the Bessel function $K_{5/3}(z)$ by
\[ F(x) = x \int_x^\infty K_{5/3}(z) {\rm d}z. \]

We are interested in the luminosity of a population of particles, so
we need to integrate $L_{\rm{1p}}(\omega)$ over suitable distributions
of energies and pitch angles. The energy distribution is that given
above, while the pitch angle distribution that we use is an isotropic
one, where the probability distribution is $p(\theta) {\rm d}\theta =
\frac{1}{2} \sin\theta\:{\rm d}\theta$. Thus, the integrated
luminosity from such a population is
\[ L(\omega) = \frac{\sqrt{3}e^3B\kappa}{16\pi^2\epsilon_0cm_e}
\int_0^\pi \sin^2\theta \left( \int_1^{\gamma_c} \gamma^{-p} F(x) \ 
{\rm d}\gamma \right) {\rm d}\theta \]

An example of such a spectrum is shown in Fig.~\ref{fig-synch}, for
$\gamma_c=10^4$ and $B=10^{-4}{\rm T}=1{\rm G}$ (the value of $\kappa$
has been taken to be 1). The peak frequency $\nu_c$ depends on these
two values, and can be shown by simple arguments \cite{blandford90} to
be approximated by $\nu_c \sim \gamma_c^2 B\ \text{MHz}$ (where $B$ is
measured in Gauss).

The slope of the power law tail (at frequencies $\nu\ll\nu_c$) is
related to the energy power law index by $\alpha_S = (p-5)/2$ (where
$f\propto\lambda^{\alpha_S}$). The energy distribution for the
spectrum in Fig.~\ref{fig-synch} is taken to be
$N(\gamma)\propto\gamma^{-2.5}$ (\ie $p=2.5$), giving a power law of
$f_\lambda \propto \lambda^{-1.25}$.

We consider here a range of $p$ values from $p=2.0$ to $p=3.0$, which
gives a range of long-wavelength power law slopes of $\alpha_S=-0.5$ to
$\alpha_S=-1.0$. This range covers the distribution of radio-to-optical
slopes observed in optical synchrotron jets
\cite{scarpa99b}. Allowing $p$ to vary does not significantly alter
the results of our analysis -- see Section~\ref{sec-discuss} for
further discussion.

We also note here that a value of $p>3$ means that the $\nu F_\nu$
flux will increase towards longer wavelengths (since $F_\nu \propto
\nu^{-(p-1)/2}$ and so $\nu F_\nu \propto \nu^{-(p-3)/2}$) and this
results in the radio flux being severely overestimated by the fitted
synchrotron component, since the radio emission always has a lower
$\nu F\nu$ flux than the optical. (This assumes that the same
synchrotron component is responsible for both the optical and radio
emission, which is an assumption commonly made, particularly for the
modelling of optical synchrotron jets \cite{meisenheimer96}.)

\subsection{Model fitting}

These two components (the blue power law and the synchrotron
component) are combined linearly to form a model $f_C(\lambda) = a
\lambda^{-1.7} + b f_{synch}(\lambda)$ that is fit to the data in the
same way as the power law model (that is, using $\chi^2$
minimisation). The reduced $\chi^2$ value (that is, $\chi^2/\nu$) for
each of the two models (combined and power law) are compared, and the
model with the lowest $\chi^2/\nu$ is chosen to be the best fit model.

In fitting the combined model, the location of the peak wavelength of
the synchrotron spectrum, $\lambda_p$, was allowed to vary. This
variation was allowed to occur over a range of rest frame wavelengths
such that the curvature of the spectrum caused by the turn-over
affected the synchrotron flux in the region of the data points (in
other words, we did not want to just be fitting the power law part of
the synchrotron spectrum). Quantitatively, we took the minimum peak
wavelength to be half a decade shorter than the $B$ band ($0.44\mu
{\rm m}$) shifted to the rest frame. We then considered 20 $\lambda_p$
values per decade (evenly spaced in $\log_{10}\lambda_p$), up to a
maximum peak of $10\mu {\rm m}$. For each of these synchrotron
functions, a best fit to the data was found, and then the best of
these was chosen, giving the best fit $\lambda_p$ value for that
source.

\section{Results for Physical models}
\label{sec-results}

\begin{figure*}
\epsfbox{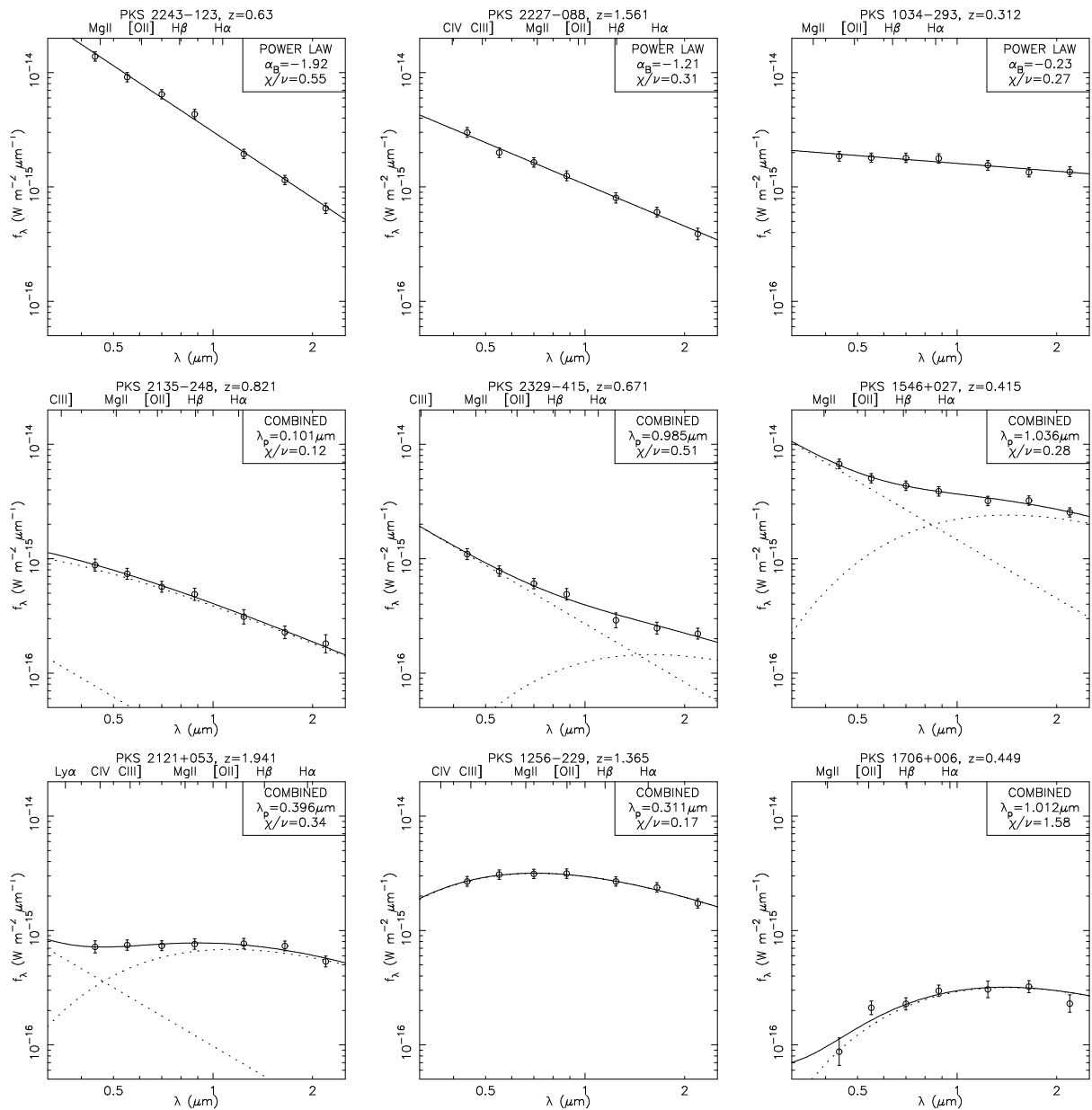}
\caption{Examples of the various types of photometry, and the fits to
them. The horizontal axis on each plot is wavelength, in $\mu {\rm
m}$, and the vertical axis is $f_\lambda$ in ${\rm W}\:{\rm m}^{-2}\:
\mu {\rm m}^{-1}$. The locations of notable emission lines are shown,
based on the redshift of each quasar. For the combined fits, the
individual components -- the power law ($\alpha_B=-1.7$) and the
synchrotron component ($p=2.5$) -- are shown as dotted lines. The
$\lambda_p$ value, where given, is for the quasar rest frame. A full
compilation of the fits to all the quasars in the sample is given in
Whiting \protect\shortcite{whiting00}.}
\label{fig-ex}
\end{figure*}

This combined model was fitted to the photometry, and compared to the
power law fits. The best fitting model was chosen on the basis of the
lowest reduced $\chi^2$ value, as described above. For the default
values of the parameters ($p=2.5$ and $\alpha_B=-1.7$), we find that
93 sources (or 79\% of the total) are well fit by one of the
models. Of these, 48 are best fit by the power law model, and 45 by
the combined model. How these numbers change with different parameter
values is discussed in Section~\ref{sec-param}. A selection of the
fits are shown in Fig.~\ref{fig-ex}, for a range of power law slopes
and synchrotron peak wavelengths.

We note here that although there are 48 sources best fit by the power
law model, many of these have combined fits that are only slightly
worse than the power law fit. This indicates that there is not a great
deal of difference between the fits of the two models. This is not the
case with the combined model sources, as for most of these the
combined model fit is a lot better than the power law fit.

\begin{figure}
\epsfbox{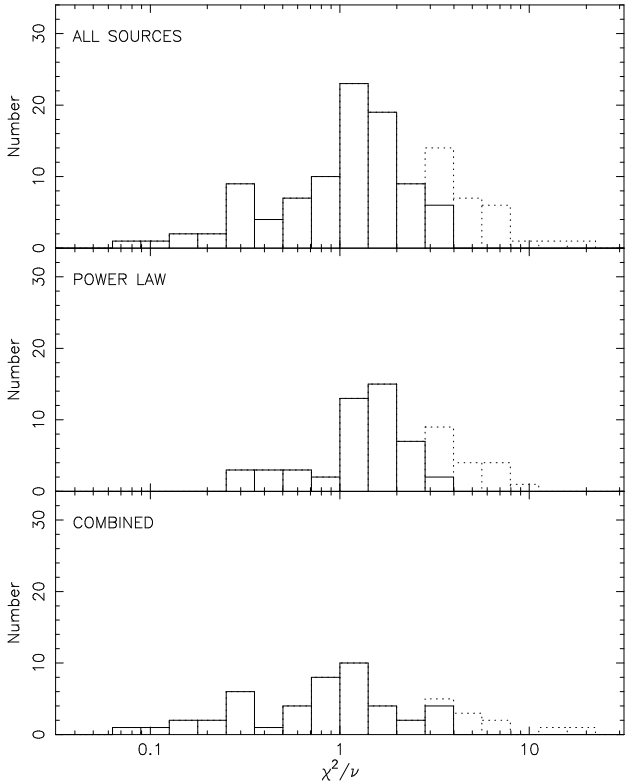}
\caption{Histograms of the reduced $\chi^2$ values, for all sources
and for each of the best-fit models. The dotted histogram shows those
sources whose fits are rejected at the 99\% confidence level.}
\label{fig-chisq}
\end{figure}

A histogram of $\chi^2/\nu$ values is shown in Fig.~\ref{fig-chisq},
with the distributions for the two different models shown
separately. The distribution for the sources best fit by the combined
model is noticeably broader than that for the power law sources, with
more sources having very low $\chi^2$ values.

\begin{figure}
\epsfbox{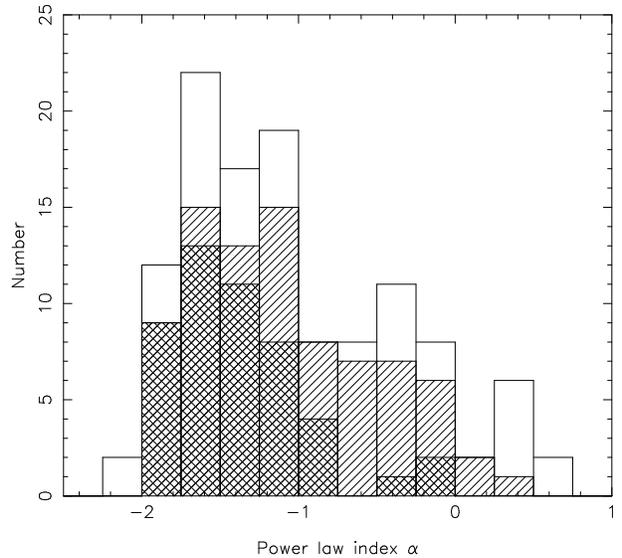}
\caption{Distributions of the fitted power index $\alpha$. The open
histogram shows the results of the power law fit to all 117 sources,
the hatched histogram shows which sources have a power law fit accepted
at 99\% confidence, while the cross-hatched histogram shows which of
those sources have power law fits better than the combined fit.}
\label{fig-ihist}
\end{figure}

While many sources have been fitted better with the combined model, a
large number are still preferentially fit with the power law. If we
plot a histogram (Fig.~\ref{fig-ihist}) of the power law indices of
those still fit by the power law model, we can see that those sources
that are preferentially fit by the power law model are the bluer
sources, while the majority of the sources with indices $\alpha>-1$
are fit better by the combined model.

\section{Fitted synchrotron components}
\label{sec-synchprops}

\begin{figure}
\epsfbox{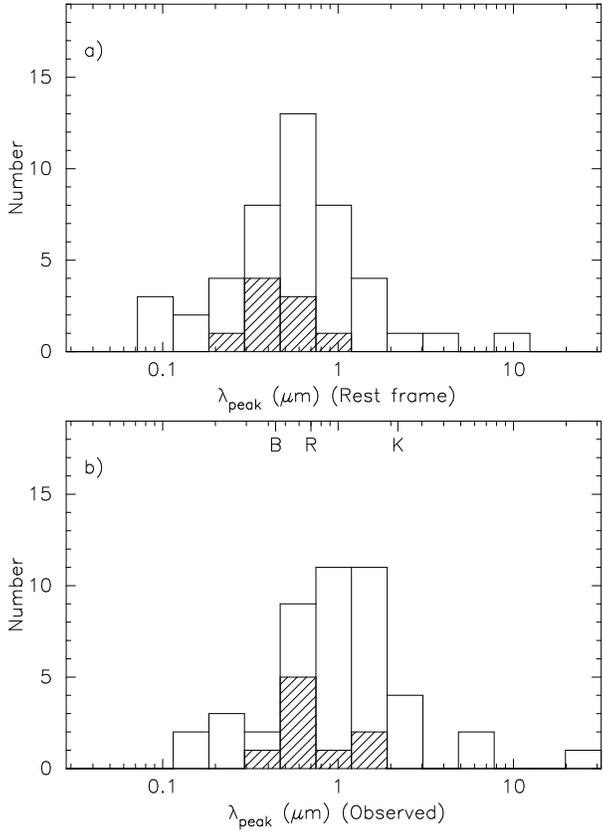}
\caption{Histogram of peak wavelengths, showing all sources fit best
with the combined model, in a) the rest frame, and b) the observed
frame. The solid histograms show all sources accepted at 99\%
confidence, while the hatched histograms show the BL Lac objects. The
locations of three prominent emission lines and photometric bands are
given for reference.}
\label{fig-peak}
\end{figure}

The properties of the synchrotron components that are fitted as part
of the combined model are of particular interest. As can be seen in
Fig~\ref{fig-peak}, the peak wavelengths are restricted to a
relatively narrow range of values (approximately a decade in
wavelength). However, this is likely to be largely a reflection of the
distribution of the wavelengths of the photometric points.

\begin{figure}
\epsfbox{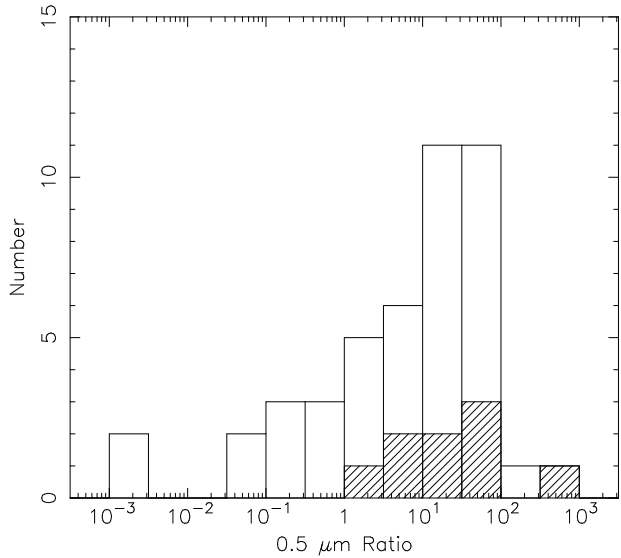}
\caption{Histogram of ratios of the synchrotron component to the power
law component of the combined model, for those sources best fit by the
combined model. The ratio is calculated at a rest-frame wavelength of
$0.5 \mu {\rm m}$. The hatched histogram shows the BL Lac objects.}
\label{fig-ratio}
\end{figure}

The strength of the fitted synchrotron component varies considerably
from source to source. In Fig.~\ref{fig-ratio}, the ratio of the
synchrotron and power law components at a rest-frame wavelength of
$0.5 \mu {\rm m}$ is shown for all sources best fit by the combined
model. The main bulk of this distribution spans nearly four orders of
magnitude. This large range of values, which is also seen in the
normalisations of the individual components, indicates that we are
seeing a continuum of variations of these components, probably due to
variations in the strengths of the inner jet and emission from the
accretion disk and/or surrounding regions.

We also note that a small number of the sources at the high-ratio end
of the distribution are faint, red sources, that are likely to be
significantly dust-reddened. They are thus fit with a dominant
synchrotron component, as the synchrotron spectrum has the approximate
form of a power law with an exponential cut-off (which is the same as
a power law with dust extinction).

\section{Testing the fits}
\label{sec-tests}

\subsection{Polarisation}
\label{sec-pol}

We have shown that a large fraction of the sources in the PHFS show
evidence for the presence of optical synchrotron emission, where the
amount of synchrotron emission present in the spectrum changes with
wavelength. How else can we test this model? One of the key features
of synchrotron radiation is its high degree of polarisation. If there
is significant synchrotron emission at optical and NIR wavelengths,
then one would expect to be able to detect a corresponding
polarisation. Indeed, this has been used as a way to confirm the
presence of synchrotron emission in optical jets (Baade
\shortcite{baade56} provided the first example of this for the jet of
M87.)

In our combined model, the synchrotron component is the only polarised
component, as we assume that the BBB component, which is essentially
emission from the accretion disk, is unpolarised ($P < 1\%$ for the
BBB \cite{antonucci98}). Thus, the amount of polarisation will depend
on the proportion of the total flux that is due to the synchrotron
emission. Furthermore, if the relative amount of synchrotron emission
changes with wavelength, as it does with the models we have fitted,
then the amount of polarisation should also change with wavelength.

Such dependencies have been investigated previously by a number of
different authors, for samples that include some sources considered
here. Wills et al.\ \shortcite{wills92} studied a large sample of
bright, flat-spectrum core-dominant quasars, measuring their optical
polarisation. An interesting result is that they found that the
fraction of quasars with $P>3\%$ in a fixed observed passband
decreased with increasing $z$, possibly indicating that the percentage
polarisation decreases towards shorter rest frame wavelengths. This
would be consistent with the presence of a synchrotron component
turning over in the optical region.

Impey \& Tapia \shortcite{impey90} present radio and optical data for a
slightly larger sample of radio-selected quasars, including optical
polarisation measurements. They find strong statistical links between
strong optical polarisation and properties such as compact radio
structure, superluminal motion and weak emission lines. They explain
this by requiring the optical emission, as well as the compact radio
emission, to be relativistically beamed.

Smith et al.\ \shortcite{smith88} obtained multicolour ($UBVRI$)
polarisation measurements of 11 highly polarised quasars, and found
that three of these exhibited decreasing polarisation toward shorter
wavelengths, which they modelled as a combination of polarised
synchrotron emission, and two unpolarised components, from the
broad-line region and the accretion disk. None of these sources,
however, are part of our sample.

\subsubsection{Optical polarisation}

\begin{figure}
\epsfbox{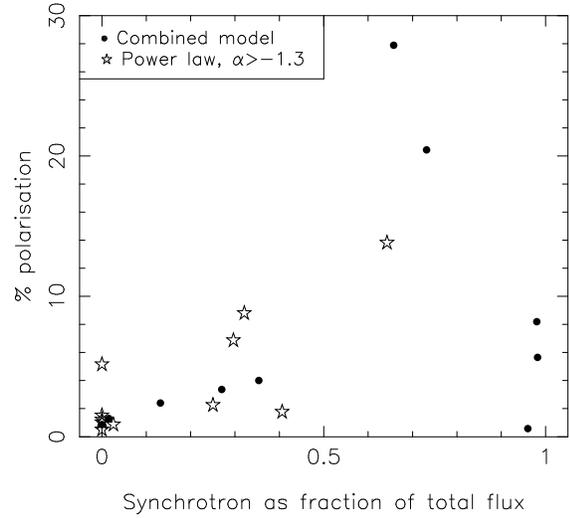}
\caption{Polarisation from Wills et al.\ \protect\shortcite{wills92}
as a function of the proportion of the total flux made up by
synchrotron. The sources are given different symbols according to the
nature of their best fit model.}
\label{fig-pol}
\end{figure}

Firstly, we wish to compare our model predictions with published
optical polarisation measurements. We use the large catalogue of
measurements compiled by Wills et al.\ \shortcite{wills92}. We want to
compare these measurements with the fractional amount of flux due to
the fitted synchrotron component. However, since the observations of
Wills et al.\ were made without a filter, we have determined the
average synchrotron fraction by integrating over the range $0.3 \mu
{\rm m}$ -- $1 \mu {\rm m}$. 

We have then plotted in Fig.~\ref{fig-pol} the percentage polarisation
as a function of this average synchrotron fraction. For those sources
that were best fit by the power law model, we have calculated the
fraction from the combined model fit, and indicated these sources by a
different symbol. All the power law sources with polarisation
measurements had relatively red power law indices (\ie $\alpha >
-1.3$).

The spread in polarisation measurements at large synchrotron fractions
is much greater than at low fractions, indicating that the high
polarisation sources generally have large amounts of synchrotron
fitted to them (at the wavelengths at which the polarisation is
measured). Additionally, all but one of the sources at zero (or
near-zero) synchrotron fraction have low optical polarisation.

The two exceptions to this picture are 0202$-$172 and
1020$-$103. Firstly, 1020$-$103 has a very high synchrotron fraction
($\sim 95\%$), but has very little optical polarisation
($P=0.58\%$). In this case, the source has a power law continuum with
a slight curvature, which is fit well by an almost pure synchrotron
curve. It may be that this curvature is due to other effects than
synchrotron, which explains the lack of polarisation. A possible
candidate is contamination from the very strong \ha line,
which would boost the continuum level in the centre of the
SED. 

Secondly, 0202$-$172 has a measured polarisation of 5.15, but also has
a very blue power law continuum ($\alpha = -1.85$). This continuum
slope is possibly too steep to attribute to synchrotron (it implies a
value of $p = 1.3$ -- in turn implying a rather flat energy
distribution ). However, the location of the synchrotron peak may have
shifted in the period since the polarisation measurements were made
(the polarisation measurements were taken 8 years prior to our
photometry observations).

\subsubsection{Near-infrared polarisation}

\begin{figure*}
\epsfbox{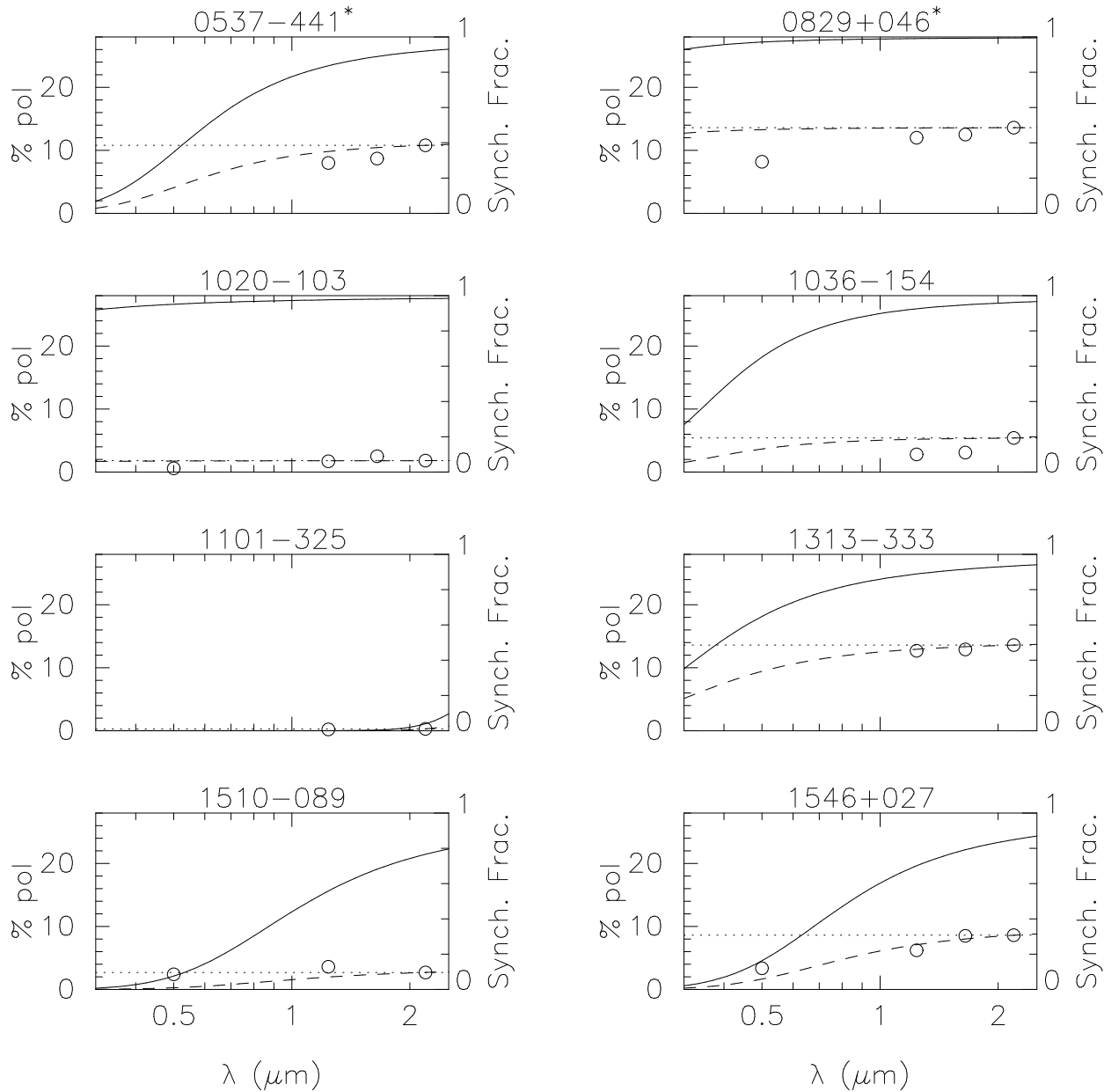}
\caption{Near-infrared polarisation of a selection of quasars as a
function of wavelength. Also shown are optical measurements from Wills
et al.\ (1992). The solid lines show the fraction of the total flux
made up by the synchrotron component (right-hand axis), while the
dashed line shows this normalised (see text) to the $K$-band data
point, and is in polarisation units (left-hand axis). The dotted line
shows constant polarisation, normalised to the $K$-band data point as
well. A $^\ast$ denotes the object is a BL Lac.}
\label{fig-mypol}
\end{figure*}

To try to avoid the problem of non-simultaneity of the photometry and
polarisation observations, we obtained polarisation measurements of 8
quasars in the NIR, using the {\it IRIS} instrument on the Anglo
Australian Telescope. The details and results of the observations will
be presented elsewhere (Whiting et al., in preparation). These
measurements are nearly simultaneous with the photometry measurements
(a difference of $\sim 40$ days), and so can be directly related to
the fitted components. The wavelength dependence of the polarisation
can then provide an important test on the models we are fitting.

Since the synchrotron component is the only polarised component, one
would expect the percentage polarisation to be directly related to the
amount of synchrotron flux present, and, in fact, the percentage
polarisation will be directly proportional to the ratio of synchrotron
flux to total flux. In Fig.~\ref{fig-mypol}, we have plotted the
polarisation of each of these quasars as a function of wavelength,
and, on the same plot, the synchrotron ratio normalised (arbitrarily,
as it is the wavelength dependence we are interested in, not the
precise normalisation) to the longest wavelength polarisation point
(which is usually the $K$-band point). For completeness, we also show
a line of constant polarisation, normalised to the $K$-band point. All
these sources are best fit with the combined model, with the exception
of 1101$-$325, which is fit best by the power law model (its SED
taking the form of a blue power law, $\lambda^{-1.81}$). We have used
the combined model fit to it for the purposes of
Fig.~\ref{fig-mypol}. 

We have also included polarisation measurements from Wills et al.\
\shortcite{wills92} where they exist. Again, these are put at an observed
wavelength of $0.5\mu {\rm m}$. Note, of course, that these points are
not simultaneous with the NIR points.

The fitted synchrotron component generally replicates well the
wavelength dependence of the polarisation, although for some sources,
such as 1313$-$333, the points are equally well given by a constant
polarisation component. This is what you would expect from a pure
synchrotron component, and these sources are typically BL Lac objects,
from which you would expect to see a synchrotron-dominated SED. A
notable exception to this is 0537$-$441, whose polarisation is not fit
well by a constant synchrotron component, but is fit better by a
combination of a synchrotron and a significant power law
component. However, it is apparent from these plots that having
simultaneous optical polarisation measurements would better help
discriminate between the combined model and a constant polarisation
model.

\subsection{Emission lines}
\label{sec-emline}

Synchrotron flux emitted from the PHFS quasars is most likely to come
from a thin, relativistic jet, and the radiation will not be
isotropic. This means that there will be very little synchrotron
radiation directed towards the emission line clouds, which have a much
larger covering angle. Thus, the ionisation of these clouds will be
due to the continuum emission from the central accretion disk region
-- in other words, the Big Blue Bump.

One might expect that adding a non-ionising synchrotron component to
the continuum will have the effect of reducing the equivalent width of
the emission lines from the BLR, due to the fact that the flux in the
emission lines is not changed but the continuum flux is increased.

To test this prediction, we compared the equivalent widths of five
emission lines (\civ\ 1549, \ciii\ 1909, \mgii\ 2798, \hb, and the
doublet \oiii\ 4959,5007) with the ratio of synchrotron to
continuum flux at the line wavelength, to see if some form of an
anti-correlation is present. The details of the observations will
presented elsewhere \cite{francis-spectra}. Objects that had spectra
taken were essentially a random sample of the PHFS (subject to
visibility during the observing run).

\begin{figure*}
\epsfbox{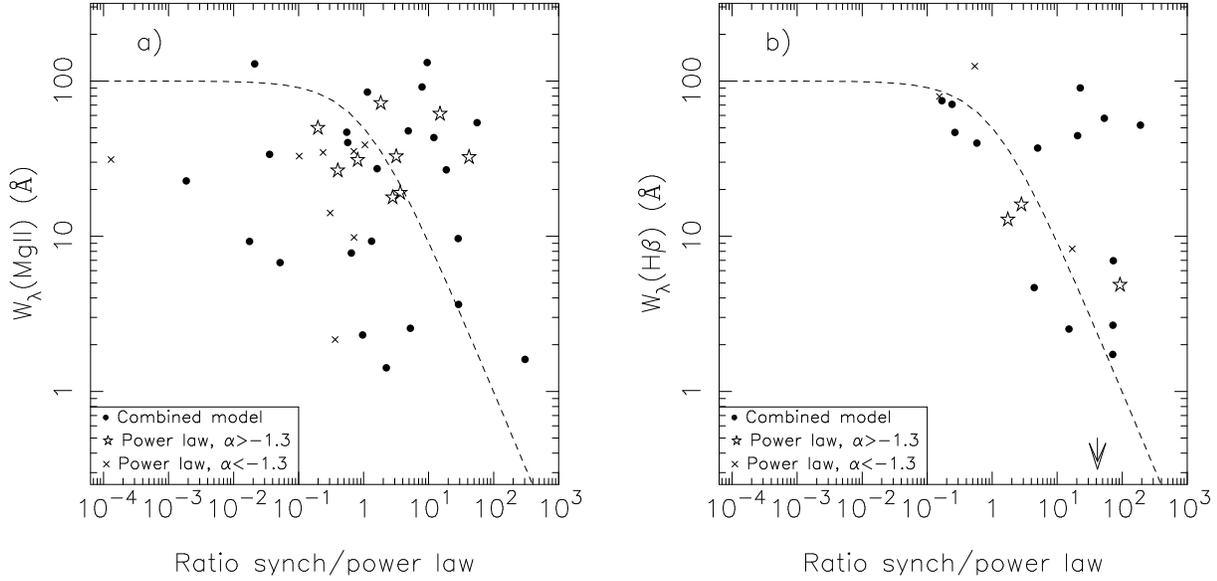}
\caption{The equivalent width of a) the \mgii\ and b) \hb\ lines as a
function of the ratio of synchrotron to continuum flux at the emitted
wavelengths. The sources are given different symbols according to the
nature of their best fit model (the arrow indicates an upper limit for
a source that was best fit with the combined model). Ratios for
sources best fit by the power law model are calculated from the
combined model fits, and so are upper limits to the ratios (for the
crosses and stars). The dashed lines represent the expected change in
equivalent width with increasing amount of synchrotron, for an
emission line with intrinsic equivalent width of 100\AA.}
\label{fig-ew}
\end{figure*}

In Fig.~\ref{fig-ew}, we show the results of this comparison for the
\mgii\ and \hb\ lines, which are the two broad lines with the longest
wavelength (and hence the two lines most likely to show a reduction in
equivalent width). We have also plotted those sources best fit with
the power law model. The value of the ratio used for these sources was
taken from fitting the combined model, and so are upper limits to the
ratio. 

The \mgii\ line does not show much relationship to the synchrotron
ratio, while the \hb\ line does show a reduction in equivalent width
with increasing amount of synchrotron. This lends some support to the
hypothesis that excess synchrotron light is present. The difference in
the two plots is likely due to the presence of the turnover in the
synchrotron flux, so that is has less effect at the shorter wavelength
of \mgii. There are, however, several sources that have high
synchrotron ratios together with high equivalent widths. The
implications of these results are discussed in
Section~\ref{sec-discuss}.

\section{Hot dust: an alternative to synchrotron?}
\label{sec-dust}

So far, we have shown that, for a number of sources, the optical--NIR
photometry is well fit with a power law plus a curved component, which
we have assumed to be the turnover of a synchrotron component. However
could another model be used instead of synchrotron? One possible
alternative is blackbody emission resulting from hot dust.

To test this model, we used a blackbody emission spectrum due to dust
at a temperature of 1750 K (the sublimation temperature characteristic
of dust grains consisting of graphite and silicates, \eg Laor \&
Draine \shortcite{laor93}) emitted in the quasars' rest frame. A blackbody
curve at this temperature would have its peak, in the quasar
rest-frame, at $1.66 \mu {\rm m}$. This blackbody spectrum was
combined with the same $\lambda^{-1.7}$ power law used in the combined
model to produce a model that was fit to the data.

The fits generated by this model were almost always worse than those
of the synchrotron model. This was due to the peak of the blackbody
occurring at much longer observed wavelengths than the $K$ band.  In
only 12 cases was the $\chi^2/\nu$ value from the dust model less than
that of the synchrotron model, and in most cases this was because the
dust model had one more degree of freedom. For each of these objects,
the SED took the form of a blue power law (one object was
$\lambda^{-1.4}$, and the rest were bluer than $\lambda^{-1.54}$) that
had a slight amount of reddening at the $H$ and $K$ bands, which was
fit by the presence of the dust blackbody curve. For the majority of
the sources, however, the hot dust model was a lot worse than the
synchrotron model, and so can not provide the peak in the optical/NIR
that is required to explain many of the observed SEDs.

A number of authors \cite[for example]{sanders89} have argued for the
existence of a near-IR bump, somewhere around $3\mu {\rm m}$,
corresponding to blackbody emission from hot dust. Sanders et al.\
\shortcite{sanders89}, also mention the presence of a local minimum at
$1\mu {\rm m}$, which they note for its ``universality''. They
interpret this to be due to the finite sublimation temperature of
dust, causing a drop in the blackbody emission, combined with the rise
in flux of thermal emission from hot ($T\gtrsim10,000$K) gas.

Any dip observed in our data occurs at wavelengths which are too short
to be attributed to hot dust. To observe a $1\mu {\rm m}$ dip, we
would need photometry at longer wavelengths.

\section{Effect of emission lines on photometry.}
\label{sec-em-effect}

As mentioned in Section~\ref{sec-data}, the presence of a strong
emission line in the wavelength range of one of the filters will
raise the SED level above that of the continuum, which will increase
the systematic error in the photometry. 

To test this effect, and to see how the fits to the photometry are
affected, we examined several sources whose spectra showed significant
emission lines. To calculate the contribution of the emission line, we
define the total flux in the line as $F_{\text line}$ and the total
continuum flux {\em under} the line as $F_{\text cont}$. Then the
change in magnitude is given by
\begin{align*}
\Delta m 	&= 2.5 \log_{10}\left(\frac{F_{\text line}+
				F_{\text cont}}{F_{\text cont}}\right)\\
		&= 2.5 \log_{10}\left
			( \frac{\Delta\lambda +	W_\lambda}
				{\Delta\lambda} \right)
\end{align*}
where $W_\lambda$ is the equivalent width of the line, and
$\Delta\lambda$ is the wavelength range over which the flux is
measured. The equivalent widths and fluxes were computed using the
{\em splot} routine in {\sc iraf}.

Note that we are only able to do this analysis for a small number of
sources, as we do not have spectra for all sources, and those spectra
that we do have are often poor quality, and in almost all cases
non-simultaneous.

The values of the change in magnitude, when considered over the same
wavelength range as that of the broad-band filters ($\Delta\lambda\sim
1000 - 2400$\AA\ for the optical bands), ranged from 0.05 -- 0.25
magnitudes.  By artificially reducing the flux in the relevant band by
this amount, we could evaluate the change in $\chi^2$ caused by the
presence of the emission line. This was done for several sources that
had strong emission lines present in their spectra (the spectra for a
large number of PHFS sources will be presented elsewhere
(Francis et al., in preparation)). A few specific examples:
\begin{itemize}
\item 1510$-$089 has a strong \ha\ line in the $I$ band ($\Delta m =
0.14$), as well as a combination of a strong \hb\ line and prominent
\feii\ emission in the $R$ band ($\Delta m = 0.11$). The removal of
flux corresponding to these lines caused a reduction in $\chi^2/\nu$ of
nearly 50\% (1.28 -- 0.68), without greatly changing the resultant
fit, although the fitted synchrotron peak was at a slightly longer
wavelength.
\item 1725$+$044, similarly, has \ha\ in $I$ band ($\Delta m = 0.13$)
and \hb\ + \oiii\ in $R$ band ($\Delta m = 0.07$), and removal of these
more than halves $\chi^2/\nu$ from 2.49 to 1.17, without changing the
location of the peak of the synchrotron.
\item 1036$-$154 has a large \mgii\ line in the $B$ band ($\Delta m =
0.17$), which causes a noticeable upturn in the SED. Removal of this
line reduces the $\chi^2/\nu$ from 0.77 to 0.56, without changing the
location of the peak wavelength, although the $0.5 \mu {\rm m}$ ratio
increases slightly (from 5.3 to 7.2). 
\item 1136$-$135 is initially fit with a pure power law, and this
remains the case after removal of the \mgii\ line ($\Delta m = 0.09$ in
$B$) and both the \hb\ and the strong \oiii\ (total $\Delta m = 0.11$ in
$I$). The power law index softens slightly (from -1.85 to -1.81) and
the $\chi^2/\nu$ value decreases from 0.43 to 0.29.
\end{itemize} 

In conclusion, by selecting quasars with strong emission lines, we
have demonstrated that the largest changes to the SED are $\Delta m
\sim 0.25$ in one wave-band. In most cases investigated, removal of
the line flux improved the $\chi^2/\nu$ value, but did not
significantly alter the nature of the fit. When good quality
long-wavelength spectra (preferably at least quasi-simultaneous with
the photometry, which is not the case here) are available for these
quasars, it should be possible to recalculate the fits, taking the
emission line contributions into account, although we do not expect
the general conclusions to change.

\section{Discussion}
\label{sec-discuss}

\subsection{Types of sources}

The fitting of a single power law to the photometry, as detailed in
Section~\ref{sec-powerlaw}, separates the sources into classes
depending on the goodness of the power law fit and the value of the
spectral index. The combined model fits have refined
this description so that we are able to talk about the different types
of sources present in the sample.

Firstly, the bluest sources are all fitted with the power law model,
and are generally consistent with being the same type of sources as
optically-selected quasars. The power law model also best fits a
number of objects of intermediate slope ($\alpha\sim-1.2$), as well as
a few red sources ($\alpha>-0.5$). These latter sources, as seen in
Section~\ref{sec-pol}, are generally high polarisation sources, and so
are likely to be synchrotron dominated (they of course show no
evidence for a turn-over similar to that being fit by the synchrotron
model, although their slope could be due to a very steep energy
distribution (\ie large $p$) or a more gradual turn-over).

\begin{figure}
\epsfbox{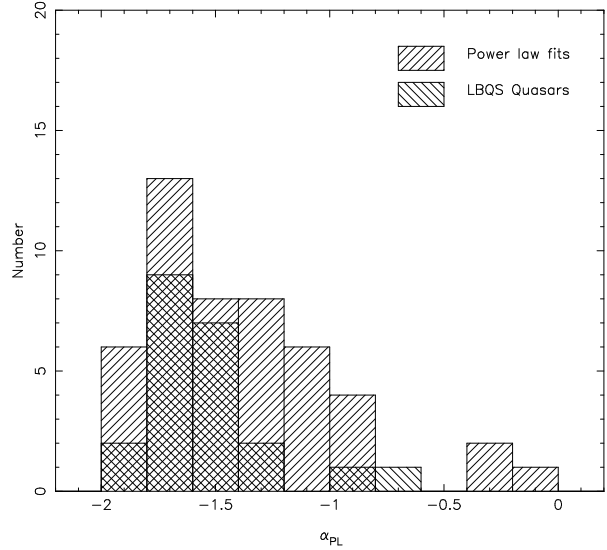}
\caption{Histograms of power law indices for the power law sources in
this paper and for the LBQS quasars from Francis
\protect\shortcite{francis96}.}
\label{fig-indexcomp}
\end{figure}

The presence of the intermediate power law slopes raises interesting
issues. Do we see similar sources in optical quasar surveys? In
Fig.~\ref{fig-indexcomp}, we have plotted the power law index
distributions for the sources best fit by the power law model, as well
as the LBQS quasars from Francis \shortcite{francis96}. While we
cannot say that the two data sets come from different parent
distributions (the Kolmogorov-Smirnov test probability for the null
hypothesis that the parent distributions are the same is 11.3\%),
there does appear to be an excess of quasars in our sample for
$-1.4\la\alpha\la-0.8$. This may indicate that the accretion disk
emission in radio loud quasars has a broader range of colours, which
would have implications for models of accretion disk emission.

The power law sources, like all sources in the PHFS, have a radio
synchrotron component. The peak of this component may be at longer
wavelengths than those covered by our observations, \ie in the
IR. This would be consistent with observations of flat-spectrum radio
quasars in the far-infrared \cite{haas98}.  These longer wavelength
peaks would be due to lower energy emitting particles, and would
therefore constitute the extension of the energy distribution, of
which we see the high energy end in the two-component sources. To test
this assertion, one needs to obtain observations at longer wavelengths
than the K band. We are in the process of obtaining observations of a
number of sources at $L$ band, and the results will be published and
discussed at a later date.

Secondly, the combined model generally best fits the redder sources,
although some intermediate sources are also fit, particularly those
with SEDs that show inflections. The synchrotron components fit to
these sources' photometry all peak in the rest-frame optical and
NIR. 

Thirdly, the other category of sources are those that are optically
faint. These sources all decrease in flux towards shorter wavelengths
(indeed, some are not detected in the $B$ band), and we suggest that
these sources are heavily dust-reddened. We note that an exponentially
reddened power law has the same form as the synchrotron component we
are fitting, and these sources are fit by a combined model with a
dominant synchrotron component. 

\begin{figure*}
\epsfbox{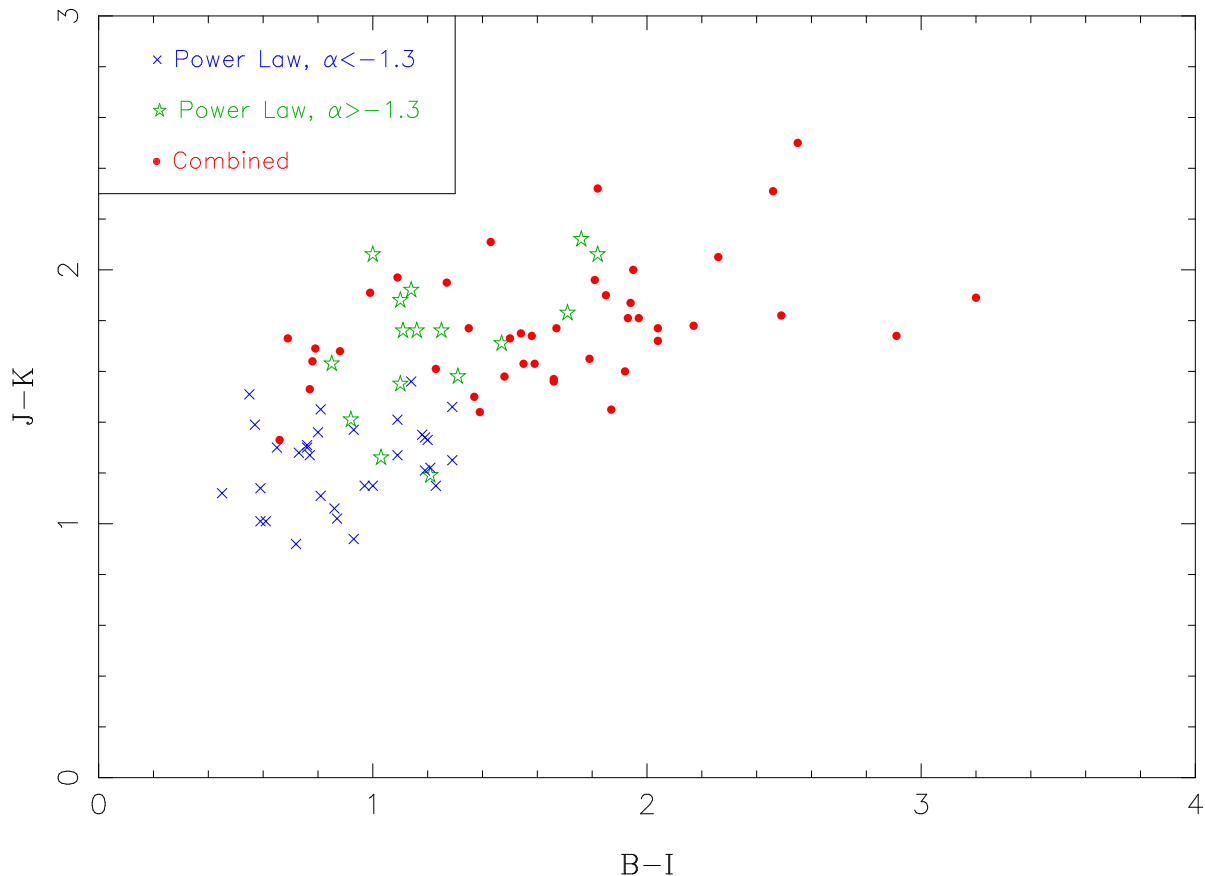}
\caption{Infrared ($J-K$) colour versus optical ($B-I$) colour for all
quasars with good model fits. The best fit model is indicated by the
symbol used.}
\label{fig-colour}
\end{figure*}

Finally, we can check the consistency of our model fits by plotting
the sources on a colour-colour diagram. In Fig.~\ref{fig-colour} we
plot the $J-K$ colours of all sources against the $B-I$ colours --
that is, the infrared colour against the optical colour. We separate
the sources into their fitted model types, separating the power law
sources by their fitted slope (using $\alpha=-1.3$ as the dividing
line). 

Clear distinctions can be made between the different model types. The
blue power law sources lie at the bottom left corner, indicating blue
colours in both optical and infrared, while as the power law becomes
redder, the sources move towards the upper-right. Many of the combined
fits are in the optically-red region of the plot, indicating that the
SED is turning over in the optical (similar to 1256$-$229 in
Fig.~\ref{fig-ex}, for example). The reddest sources in $B-I$ are
sources with optical continua that drop towards the blue, in the
manner of 1706$+$006 (see Fig.~\ref{fig-ex}), which are typical of
dusty sources, or sources dominated by host galaxy emission, and are
fit by a dominant synchrotron component. Note that 1706$+$006
is the faintest of the sources shown in Fig.~\ref{fig-ex}, and so is a
good candidate for dust extinction. Other combined sources, however,
are among the bluer sources in the optical, but have redder NIR
colours than the power law sources. These are the sources that show an
inflection, similar to 1546$+$067 in Fig.~\ref{fig-ex}, where the
synchrotron component is dominant in the NIR but turns over and has
less effect in the optical.

\subsection{Sources excluded from the sample}

As explained in Section~\ref{sec-data}, there are a number of sources
that were excluded from the sample. Some of these were excluded for
the sole reason that they did not have a measured redshift. These
sources were generally quite red and optically faint (thus explaining
their undetermined redshift), with the exception of the BL Lac object
0048$-$097, which has a power law SED. We fitted our models to the
observed wavelengths of these objects, to investigate the bias created
by excluding them. Note that this fitting does not take into account
the change in shape of the synchrotron spectrum due to the redshift of
the source. The results are summarised in Table~\ref{tab-excluded}.

\begin{table*}
\centering
\begin{minipage}{150mm}
\caption{Fitting results for sources that were excluded from the
sample purely due to their undetermined redshift. The best fit model
is given along with either the peak synchrotron wavelength (for
combined fits) or the power law index (for power law fits). $F_{0.5}$
denotes the fraction of the total flux made up by the synchrotron
component, at an observed wavelength of $0.5\mu {\rm m}$.  Note that a
$^\ast$ denotes a $V-K$ measurement, not $B-K$, since a $B$
measurement was not made (due to the faintness of the source in $B$).}
\label{tab-excluded}
\begin{tabular}{lccccccc}
Source name	&$B$ (mag)&$B-K$&Best fit model	&$\alpha$
&$\lambda_p(\mu {\rm m})$  &$F_{0.5}$	&good/bad fit\\\hline
0048$-$097	&16.12	&3.61	&Combined	&--	&0.15	&0.97	&good\\
1110$-$217	&24.41	&7.67	&Combined	&--	&1.85	&0.88	&good\\
1156$-$094	&21.95	&5.08	&Combined	&--	&1.04	&0.69	&good\\
1648$+$015	&21.87	&5.31	&Combined	&--	&1.17	&0.82	&bad\\
1732$+$094	&$B>23.5$, $V=21.15$	&$>7.13$ &Power law
						&$-0.39$ &--	&0.00	&bad\\
2056$-$369	&$V>23.5$, $R=23.45$	&$>5.39^\ast$ 
				&Power law	&0.59	&--	&0.00	&good\\
2245$+$029	&21.71	&6.22	&Combined	&--	&1.85	&0.41	&good\\
2337$-$334	&22.93	&6.54	&Power law	&0.36	&--	&0.00	&good\\
2344$-$192	&23.52	&6.27	&Combined	&--	&1.65	&0.83	&bad
\end{tabular}
\end{minipage}
\end{table*}

As can be seen, all but three of the sources are fit with the combined
model, and the three that aren't are fit with quite red power laws.
The location of the peak wavelengths are generally into the NIR, which
are longer than the bulk of the distribution of the sources with
redshifts (see Fig.~\ref{fig-peak}). Also, the synchrotron component
that is fitted is generally fairly dominant, as evidenced by the
fraction values $F_{0.5}$. However, it is unlikely that many of these
sources would truly be synchrotron sources as they show more the
characteristic shape of dust absorption and are quite faint in the
optical (again, with the exception of 0048$-$097, which was fit with
the power law tail of a dominant synchrotron model). We propose that
these sources are dust-dominated rather than synchrotron-dominated. 

\subsection{Sources without good fits}

Up to this point, we have only discussed the results for the sources
with good (\ie acceptable at the 99\% confidence level) fits. A total
of 24 sources (or 21 per cent of the total) are not fit well by either
the power law model or the combined model. What sort of sources are
these?

A few sources ($\sim 8$) have roughly power law SEDs, but with a
little curvature (`n'-shaped) in the blue end of the optical. This may
be indicative of a small amount of dust attenuation or extinction.

Most of the other sources have one or more photometric points that do
not smoothly connect with the rest of the SED. It is possible that,
for these points, at least one of the systematic errors discussed in
Section~\ref{sec-errphot} is dominating, over and above the level we
assigned. In some sources, here may also be further emission processes
present that we have not modelled.

\subsection{Alternative values of $\bmath{p}$ and $\bmath{\alpha_B}$}
\label{sec-param}

Throughout this paper, we have used values for the electron energy
index of $p=2.5$, and for the BBB spectral index of
$\alpha_B=-1.7$. We consider here the effect that changing these
values has on the results. In Table~\ref{tab-param}, we list the
numbers of sources best fit by each of the two models, for each set of
parameters, as well as the total number of sources fit by one of the
models. 

\begin{table*}
\centering
\begin{minipage}{150mm}
\caption{Number of sources fitted best by each model. Only those fits
accepted at the 99\% confidence level are shown.  Values of the
parameters used are $p=2.0$, $p=2.5$ and $p=3.0$, as well as
$\alpha_B=-1.4$, $\alpha_B=-1.7$ and $\alpha_B=-2.0$.}
\label{tab-param}
\begin{tabular}{lccccccccccc}
$p$    		&$2.0$  &$2.0$  &$2.0$  &&$2.5$	&$2.5$	&$2.5$  &&$3.0$	&$3.0$  &$3.0$  \\
$\alpha_B$	&$-1.4$ &$-1.7$ &$-2.0$ &&$-1.4$&$-1.7$	&$-2.0$ &&$-1.4$&$-1.7$ &$-2.0$ \\ \hline
Power law	&46     &44     &45     &&49	&48	&48     &&53   	&52     &52     \\
Combined	&49     &51     &50     &&44	&45	&44     &&39   	&40     &40     \\
Total		&95     &95     &95     &&93 	&93	&92     &&92   	&92     &92     \\
\end{tabular}
\end{minipage}
\end{table*}

Reducing the value of $p$ means that the synchrotron spectrum has a
bluer slope, which enhances the effect of the turn-over. Using a lower
value of $p$ in the combined model results in some sources, otherwise
fit by the power law, being instead fit by the combined model. This is
indeed seen in Table~\ref{tab-param}, where the number of sources
fit by a power law decreases as you move from $p=3$ to $p=2.5$ to
$p=2$ (if the value of $\alpha_B$ is kept constant), while the number
of sources fit by the combined model increases. Note also that more
sources have good fits for the lower values of $p$.

The $\lambda_p$ distributions are also affected by a changing $p$
value. For lower $p$ values, there are more sources with shorter
$\lambda_p$ values (that is, close to $0.1 \mu {\rm m}$). This is due
to the slope of the power law tail of the synchrotron spectrum: for
higher $p$ values the slope is redder, and so such a synchrotron spectrum
peaking around $0.1 \mu {\rm m}$ would contribute too much at the
longer wavelengths. 

For the spectral index of the blue optical power law used in the
combined model, we considered both steeper and flatter values than the
one mentioned in Section~\ref{sec-bbb}.  The dispersion found by
Francis \shortcite{francis96} for the slopes of LBQS quasars is
$\pm\sim0.3$, and so we consider here slopes in $f_\lambda$ of
$\alpha_B=-1.4$ and $\alpha_B=-2.0$.

Changing the slope of this power law (while keeping $p$ constant) has
a much less drastic effect than changing the value of $p$. As the
slope becomes steeper in $f_\lambda$ (that is, bluer), we see that
only a couple of sources fit by the power law change to be fit by the
combined model. The distribution of $\lambda_p$ values is not changed
considerably by the variation of the $\alpha_B$ values.

In summary, reasonable changes to the fiducial values of $\alpha_B$
and $p$ make little difference to the fits, both in the $\lambda_p$
distribution and the numbers fitted by each model, although the
difference will be accentuated if the values of $\alpha_B$ and $p$ are
both taken to extremes.

\subsection{Nature of synchrotron model}

Thus far, we have been considering a synchrotron model that peaks at
some $\lambda_p$ and then turns over sharply (\ie exponentially). The
reasons we chose this were for consistency with other modelling done
for optical synchrotron emission \cite{meisenheimer96}, and to
provide the maximum contrast with the power law model. However,
synchrotron emission could alternatively be present in the form of
a power law. What if we have a model of the same form as the combined
model, but with this power law synchrotron model instead? Is this any
better at fitting the observations?

We constructed such a model, being a linear combination of the blue
(BBB) power law from earlier, and a power law of variable index:
$f(\lambda) = a\lambda^{-1.7} + b\lambda^{\alpha_S}$. To distinguish
it from the BBB power law, we restricted the indices to the range
$-1.6 < \alpha_S < 1.5$. This model was then fitted to the photometry
in the same way as previously. Note that the special case of $a=0$ is
simply the power law fit from Section~\ref{sec-powerlaw} (with the
index restricted to lie in the above range). 

\begin{figure}
\epsfbox{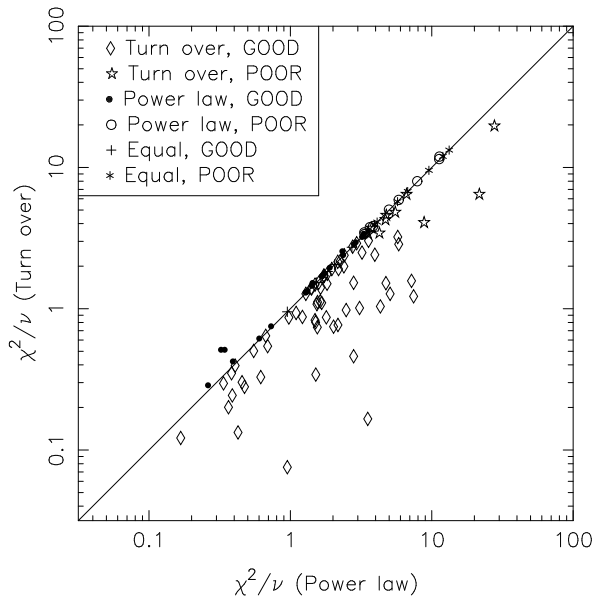}
\caption{Values of $\chi^2/\nu$ for the combined models featuring
either power law synchrotron models or synchrotron models with a
turn-over. The symbols relate to which model is preferred, and whether
the fit is accepted at the 99\% confidence limit or not (\ie good or
poor). The line shown is the line of equal $\chi^2/\nu$.}
\label{fig-2pl}
\end{figure}

The $\chi^2/\nu$ values for this model are plotted in
Fig.~\ref{fig-2pl} against those for the combined model detailed in
Section~\ref{sec-models}. The symbols used indicate which model is
preferred, and whether it is a good or bad fit. For sources that are
bluer than $\lambda^{-1.7}$, the best fit is always with no
synchrotron component (for both forms of the synchrotron model), as
the presence of the second component will redden the blue power
law. Hence, for these sources, both the original combined model and
the two power law model give the same fit, consisting of just the
$\lambda^{-1.7}$ (\ie BBB) power law. These sources are indicated by
the ``Equal fit'' symbols on Fig.~\ref{fig-2pl}.

The power law synchrotron model obviously does not do as well at
explaining the optical/NIR SEDs of the quasars as the model with a
turn-over. Only 22 sources have a good power law synchrotron fit that
is better than their fit from the synchrotron model with a turn-over,
and only two of these (1034$-$293 and 2329$-$415, with indices of
$\alpha_S=-0.21$ and $\alpha_S = 0$ respectively) are significantly
better. These power law synchrotron sources may have turn-overs at
shorter wavelengths, beyond our currently available data. The only way
to tell would be to obtain UV photometry (ideally contemporaneous with
the optical/NIR). 

We note that many of the sources from Fig.~\ref{fig-ihist} that are in
the cross-hatched histogram (that is, had better single power law fits
than combined fits) and had $\alpha > -1.6$, also had fits with the
two power law model that were better than those with the turnover
synchrotron model. (This is not always the case, however, since going
from one power law to two decreases the degrees of freedom, so the
$\chi^2/\nu$ value will be greater.) In most of these cases, the blue
power law (the $\lambda^{-1.7}$ component) is either absent or has a
very low normalisation.

For most sources, however, the synchrotron model with a turn-over
gives a much better fit to the observed optical/NIR SEDs (many of
these sources show large ratios of synchrotron to blue power law
flux). This result provides clear evidence that there is a break or a
cutoff present in the energy distribution of the radiating particles.

\subsection{Emission lines and BL Lac objects}

We can use the results of the emission line study in
Section~\ref{sec-emline} to draw some conclusions about BL Lac
objects, which are noted for having very small or non-existent
emission lines. This is usually explained by the presence of a beamed
synchrotron component from a jet directed towards us \cite[for
example]{blandford78}. The results shown in Fig.~\ref{fig-ew} seem to
support this explanation, as the equivalent widths tend to decrease
with increasing amount of fitted synchrotron, as you would expect if
there was an extra synchrotron component over the top of the emission
lines.

Also, the fits to the BL Lacs in our sample are suggestive of strong
synchrotron components. All nine BL Lacs are well fit by the combined
model, and the $0.5 \mu {\rm m}$ ratios of the synchrotron to power
law components range from 3 (0537$-$441) to over 700 (1256$-$229) (see
Fig.~\ref{fig-ratio}). These ratios, on average, are higher than the
rest of the population (although it is only a small group).

However, there may be more to the story. As can be seen from
Fig.~\ref{fig-ratio}, there are a number of sources with comparable
ratios to the BL Lacs. Also, Fig.~\ref{fig-ew} shows signs of a large
dispersion in the equivalent width distribution, particularly at the
high-ratio end, with sources present that have relatively high ratios
at the emission lines in question, as well as large equivalent
widths. Perhaps there are other factors that determine whether an
object is a BL Lac or not. One factor that is likely to be important
is the strength or dominance of the emission line region, with BL Lacs
having an intrinsically weak emission line flux.  Such an idea agrees
with other studies: for example, Ghisellini et al.\
\shortcite{ghisellini93}, who found that BL Lac objects had
significantly lower Doppler factors than core-dominated quasars. This
was interpreted as suggesting that BL Lac objects have intrinsically
weaker emission lines than core-dominated quasars.

\section{Conclusions}

The results of this paper can be summarised in the following conclusions:

\begin{enumerate}

\item Radio-loud quasars require one of at least two components to fit
the rest frame optical SEDs. These two components are well modelled by
an optical power law which in some sources is similar to the blue
power law seen in optically-selected quasars, and a synchrotron
component that turns over in the range $0.1 \mu{\rm m} \la \lambda_p
\la 3 \mu{\rm m}$.

\item The model fits require the synchrotron component to have a break
or turn-over at some wavelength $\lambda_p$ in the observed wavelength
range. A synchrotron spectrum with this break gives a much better fit
to the data than a pure power law synchrotron model. The value of
$\lambda_p$ provides strong constraints on the modelling of particle
acceleration and emission mechanisms.

\item There is more than 4 orders of magnitude variation in the ratio
of the synchrotron component to the power law component in those
sources fit by the combined model.  All the sources identified as BL
Lacs lie at the high end of the distribution.

\item The fitting procedure is robust to reasonable changes in the
values of the model parameters.  The fits are statistically consistent
with the data.

\item There may be an excess of red power law sources compared to
optically-selected quasars. This will have implications for emission
models of accretion disks.

\item Sources which are well-fitted by power laws have a synchrotron
component which may peak in the IR. In the optical, these sources may
be similar to optically-selected quasars.

\item Optical polarisation measurements (taken from the literature)
support the synchrotron model fits to the SEDs.

\item There is some evidence for the equivalent widths of emission
lines being reduced by the presence of the synchrotron component. This
supports the hypothesis that BL Lac objects have a dominant
synchrotron component, although other effects may contribute to the
lack of emission lines in BL Lacs.

\item The red component fitted to the data, which has a turnover at
some peak wavelength, cannot be hot dust.  Emission from hot dust
cannot peak at $\lambda_{\text{rest}} < 1.66 \mu {\rm m}$ as is
required by our data.

\item The red sources excluded from our analysis may provide
interesting examples of extreme sources in the sample, \eg very dusty
sources, high redshift sources etc.

\end{enumerate}

\section*{Acknowledgements}

MTW acknowledges the generous help of the Grimwade Scholarship from
the University of Melbourne for assistance in carrying out this
research. Thanks are due to Frank Masci, for use of some of the
polarisation data prior to publication, and to the referee (Patrick
Leahy) for some very useful comments which improved the paper.

\end{document}